\documentclass{acm_proc_article-sp}

 \usepackage{hyperref}
\usepackage[ruled,vlined,nofillcomment]{algorithm2e}
\usepackage{graphicx}
\usepackage{balance}  
\usepackage{amsfonts}
\usepackage{array}
\usepackage{url}
\usepackage{multirow}
\usepackage{placeins}
\usepackage{subfig}
\usepackage{tablefootnote}
\usepackage{amssymb}
\usepackage{amsmath}

\usepackage{xspace}
\usepackage{pgfplots}

\newtheorem{definition}{Definition}

\begin{document}


\title{Big Graph Mining: Frameworks and Techniques}

%
%
%
%
%

%
%
%
%
%
%

\numberofauthors{2}

\author{
\alignauthor
Sabeur Aridhi \\ 
       \affaddr{Aalto University, School of Science, P.O. Box 12200, FI-00076, Finland. }\\
       \email{sabeur.aridhi@aalto.fi}
\alignauthor 
Engelbert Mephu Nguifo\\
     \affaddr{LIMOS - UBP - Clermont University, BP 10125, 63173, Clermont Ferrand, France.}\\
       \email{mephu@isima.fr}
}

\maketitle
\balance
\sloppy
\begin{abstract}
Big graph mining is an important research area and it has attracted considerable attention. 
It allows to process, analyze, and extract meaningful information from large amounts of graph data. 
Big graph mining has been highly motivated not only by the tremendously increasing size of graphs but also by its huge number of applications. 
Such applications include bioinformatics, chemoinformatics and social networks. One of the most challenging tasks in big graph mining is pattern mining in big graphs. 
This task consists on using data mining algorithms to discover interesting, unexpected and useful patterns in large amounts of graph data. 
It aims also to provide deeper understanding of graph data. 
In this context, several graph processing frameworks and scaling data mining/pattern mining techniques have been proposed to deal with very big graphs. 
This paper gives an overview of existing data mining and graph processing frameworks that deal with very big graphs. 
Then it presents a survey of current researches in the field of data mining / pattern mining in big graphs and discusses the main research issues related to this field. 
It also gives a categorization of both distributed data mining and machine learning techniques, graph processing frameworks and large scale pattern mining approaches.
\keywords{Big graphs; data mining; pattern mining; graph processing frameworks}\\
\end{abstract}




\section{Introduction}
Over the last decade, big graph mining has attracted considerable attention. 
This field has been highly motivated, not only by the increasing size of graph data, but also by its huge number of applications.  
Such applications include the analysis of social networks~\cite{5992567,Seidman1983269}, Web graphs~\cite{Alvarez}, as well as spatial networks~\cite{JTLU23}. 
It has emerged as a hot topic that consists on the deliver of deeper understanding of the graph data. 
Frequent pattern mining is a main task in this context and it has attracted much interest. 
Several algorithms exist for frequent pattern mining. 
However, they are mainly used on centralized computing systems and evaluated on relatively small databases \cite{Worlein:2005:QCS:2101235.2101276}.
Yet, modern graphs are growing dramatically which makes the above cited approaches face the scalability issue. 
Consequently, several parallel and distributed solutions have been proposed to solve this problem \cite{Malewicz_et_al} \cite{Tatikonda_et_al} \cite{Bahmani:2012:DSS:2140436.2140442} \cite{Liu:2009:MPF:1616625.1616658} \cite{Hill:2012:IMA:2382936.2383055} \cite{Luo:2011:TES:1996686.1996690}. 
In addition to that, many distributed frameworks have been used to deal with the existing deluge of data. 
These distributed frameworks abstract away most of the challenges of building a distributed system and offer simple programming models for data analysis \cite{Jin201559}. 
Most of them are quite simple, easy to use and able to cope with potentially unlimited datasets. 

In this paper, we first study existing works on the field of big data analytics. Thus, we present a survey on distributed data mining and machine learning approaches. 
Then, we study existing graph processing frameworks and we highlight pattern mining solutions in big graphs.  
With reference to the literature we can identify many different types of distributed graph mining techniques, with respect to the format of the input data, to produce many different kinds of patterns. 
We also give a categorization of both techniques for big data analytics, graph processing frameworks and large scale pattern mining approaches. 
Techniques for big data analytics are described according to their related programming model and the supported programming language. 
Graph processing frameworks are described according to their related programming model, the type of resources used by each framework and whether the framework allows asynchronous execution or not. 
Pattern mining approaches are described according to the input, the output of each approach and the used programming model. 

The remainder of the paper is organized as follows. 
In the following section, we present existing works on big data analytics. 
In Section 3, we present an overview of graph processing frameworks and graph processing related approaches. 
Specifically, we present works that deal with pattern mining techniques in big graphs. 
Finally, we discuss the presented approaches in Section 4. 

\section{Big Data Analytics}
In  this  section,  we  review  related  works  on MapReduce and distributed data mining and machine learning techniques in the context of Big Data. 
\subsection{MapReduce}
MapReduce \cite{mapreduce} is a framework for processing highly distributable problems across huge datasets using a large number of computers. It was developed within Google as a mechanism for processing large amounts of raw data, for example, crawled documents or web request logs. This data is so large, it must be distributed across thousands of machines in order to be processed in a reasonable amount of time. This distribution implies parallel computing since the same computations are performed on each CPU, but with a different dataset. MapReduce is an abstraction that allows to perform simple computations while hiding the details of parallelization, data distribution, load balancing and fault tolerance.
The central features of the MapReduce framework are two functions, written by a user: Map and Reduce. 
The Map function takes as input a pair and produces a set of intermediate key-value pairs. 
The MapReduce library groups together all intermediate values associated with the same intermediate key and passes them to the Reduce function. 
The Reduce function accepts an intermediate key and a set of values for that key. It merges these values together to form a possible smaller set of values. 

Hadoop is the open-source implementation of MapReduce. 
It is composed of two components. The first component is the Hadoop Distributed File System (HDFS) for data storage. 
The second one is the wide spread MapReduce programming paradigm \cite{mapreduce}. Hadoop provides a transparent framework for both reliability and data transfers. 
It is the cornerstone of numerous systems which define a whole ecosystem around it. 
This ecosystem consists of several packages that runs on top of Hadoop including PIG \cite{pig},  a high level language for Hadoop, HBase \cite{hbasebook},  a column-oriented
data storage on top of Hadoop, and Hive \cite{hive}, a framework for querying and managing large datasets residing in distributed storage using a SQL-like language called HiveQL. 


\subsection{Distributed machine learning and data mining techniques}
Data mining and machine learning hold a vast scope of using the various aspects of Big Data technologies for scaling existing algorithms and solving some of the related challenges \cite{AlJarrah201587}. 
In the following, we present existing works on distributed machine learning and data mining techniques. 
\subsubsection{NIMBLE}
NIMBLE \cite{nimble} is a portable infrastructure that has been specifically designed to enable the implementation of parallel machine learning (ML) and data mining (DM) algorithms. 
The NIMBLE approach allows to compose parallel ML-DM algorithms using reusable (serial and parallel) building blocks that can be efficiently executed using MapReduce and other parallel programming models.
The programming abstractions of NIMBLE have been designed with the intention of parallelizing ML-DM computations and allow users to specify data parallel, iterative, task parallel, and even pipelined computations. 
The NIMBLE approach has been used to implement some popular data mining algorithms such as $k$-Means Clustering and Pattern Growth-based Frequent Itemset Mining, $k$-Nearest Neighbors, Random Decision Trees, and RBRP-based Outlier Detection algorithm. 
\begin{figure}[t]
\centering
 \includegraphics[width=0.45\textwidth]{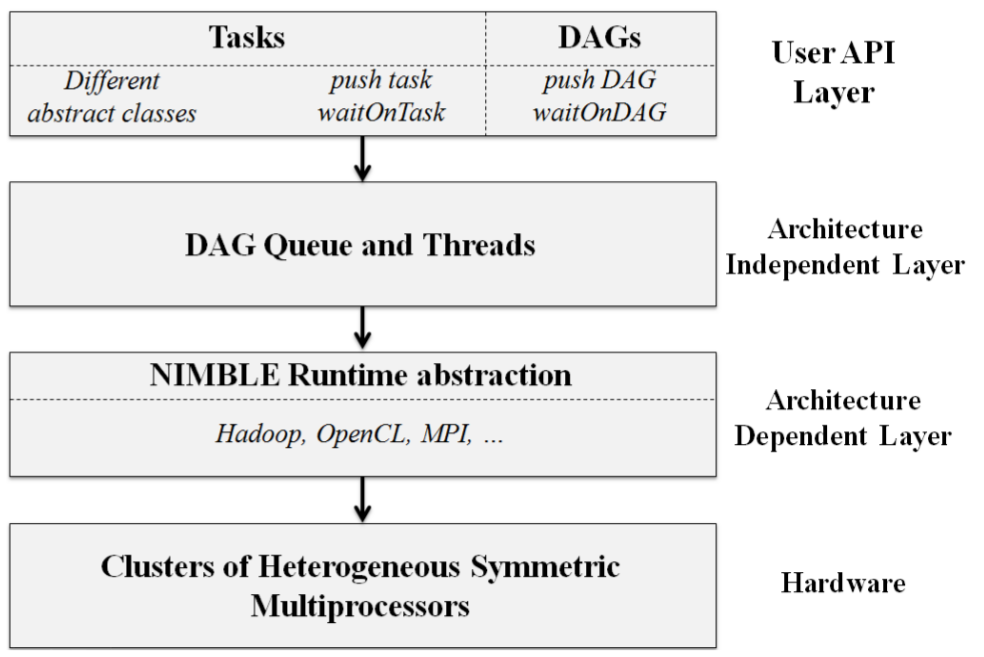}

\caption{An overview of the software architecture of NIMBLE.}
\label{nimble}
\end{figure} 
As shown in Figure \ref{nimble}, NIMBLE is organized into four distinct layers: 
\begin{enumerate}
 \item The user API layer, which provides the programming interface to the users. Within this layer, users are able to design tasks and Directed Acyclic Graphs
(DAGs) of tasks to express dependencies between tasks. A task processes one or more datasets in parallel and produces one or more datasets as output. 
\item The architecture independent layer, which acts as the middleware between the user specified tasks/DAGs, and the underlying architecture dependent layer. 
This layer is responsible for the scheduling of tasks, and delivering the results to the users. 
\item The architecture dependent layer, which consists of harnesses that allow NIMBLE to run portably on various platforms. 
Currently, NIMBLE only supports execution on the Hadoop framework. 
\item The hardware layer, which consists of the used cluster. 
\end{enumerate}
\subsubsection{SystemML}
SystemML \cite{Ghoting:2011:SDM:2004686.2005625} is a system that enables the development of large scale machine learning algorithms. 
\begin{figure}
\centering

 \includegraphics[width=0.45\textwidth]{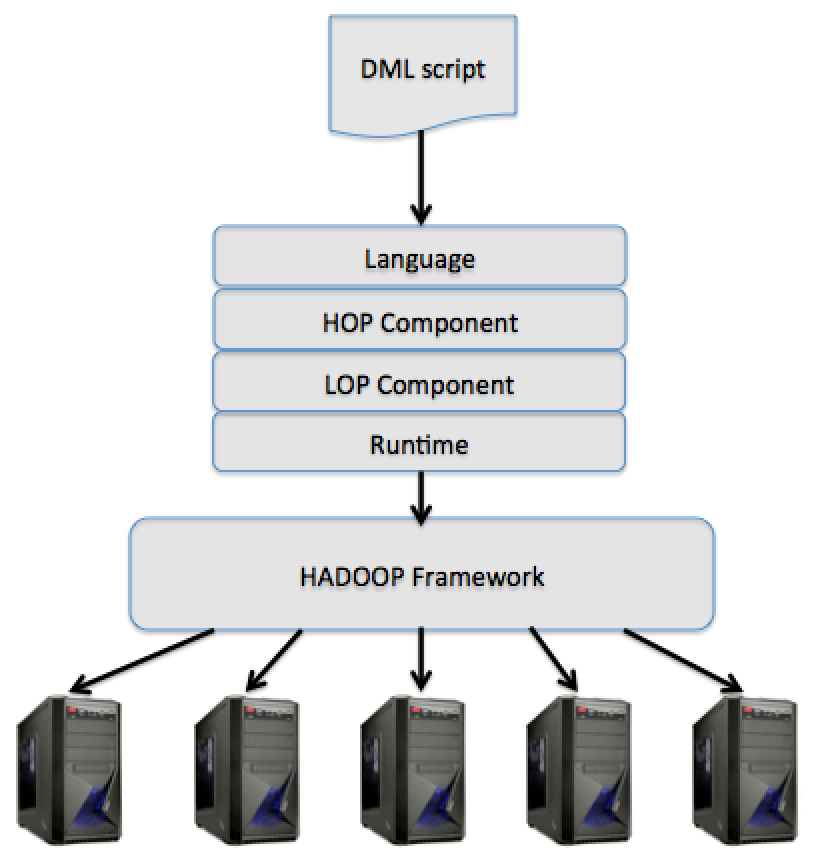}

\caption{An overview of the software architecture of SystemML.}
\label{systemml}
\end{figure} 
It first expresses a machine learning algorithm in a higher-level language called Declarative Machine learning Language (DML). 
Then, it executes the algorithm in a MapReduce environment. 
This DML language exposes arithmetical and linear algebra primitives on matrices that are natural to express a large class of machine learning algorithms. 
As shown in Figure \ref{systemml}, SystemML is organized into four distinct layers: 
\begin{itemize}
\item The Language component: It consists of user-defined algorithms written in DML.
\item The High-Level Operator Component (HOP): It analyzes all the operations within a statement block and chooses from multiple high-level execution
plans. A plan is represented in a DAG of basic operations (called hops) over matrices and scalars. 
\item The Low-Level Operator Component (LOP): It translates the high-level execution plans provided by the HOP component into low-level physical plans on MapReduce.
\item The runtime component: It executes the low-level plans obtained from the LOP component on Hadoop.
\end{itemize}
\subsubsection{Mahout}
The Apache’s Mahout project \cite{mahout} provides a library of machine learning implementations. 
The primary goal of Mahout is to create scalable machine-learning algorithms that are free to use under the Apache license. 
It contains implementations for clustering, categorization, and evolutionary programming and it uses the Apache Hadoop library to scale effectively in the cloud. 
In addition, Mahout uses the Apache Hadoop library to scale effectively in the cloud. Mahout's primary features are: 
\begin{itemize}
 \item Scalable machine learning libraries. The core libraries of Mahout are highly optimized for good performance and for non-distributed algorithms.
 \item Several MapReduce enabled clustering implementations, including k-Means, fuzzy k-Means and Mean-Shift. 
 \item Distributed Naive Bayes implementation and distributed Principal Components Analysis (PCA) techniques for dimensionality reduction. 
\end{itemize}

\subsubsection{PARMA}
PARMA \cite{parma} is a parallel technique for mining frequent itemsets (FI's) and association rules (AR's). 
PARMA is built on top of MapReduce. 

\begin{figure}[t]
\centering

 \includegraphics[width=0.45\textwidth]{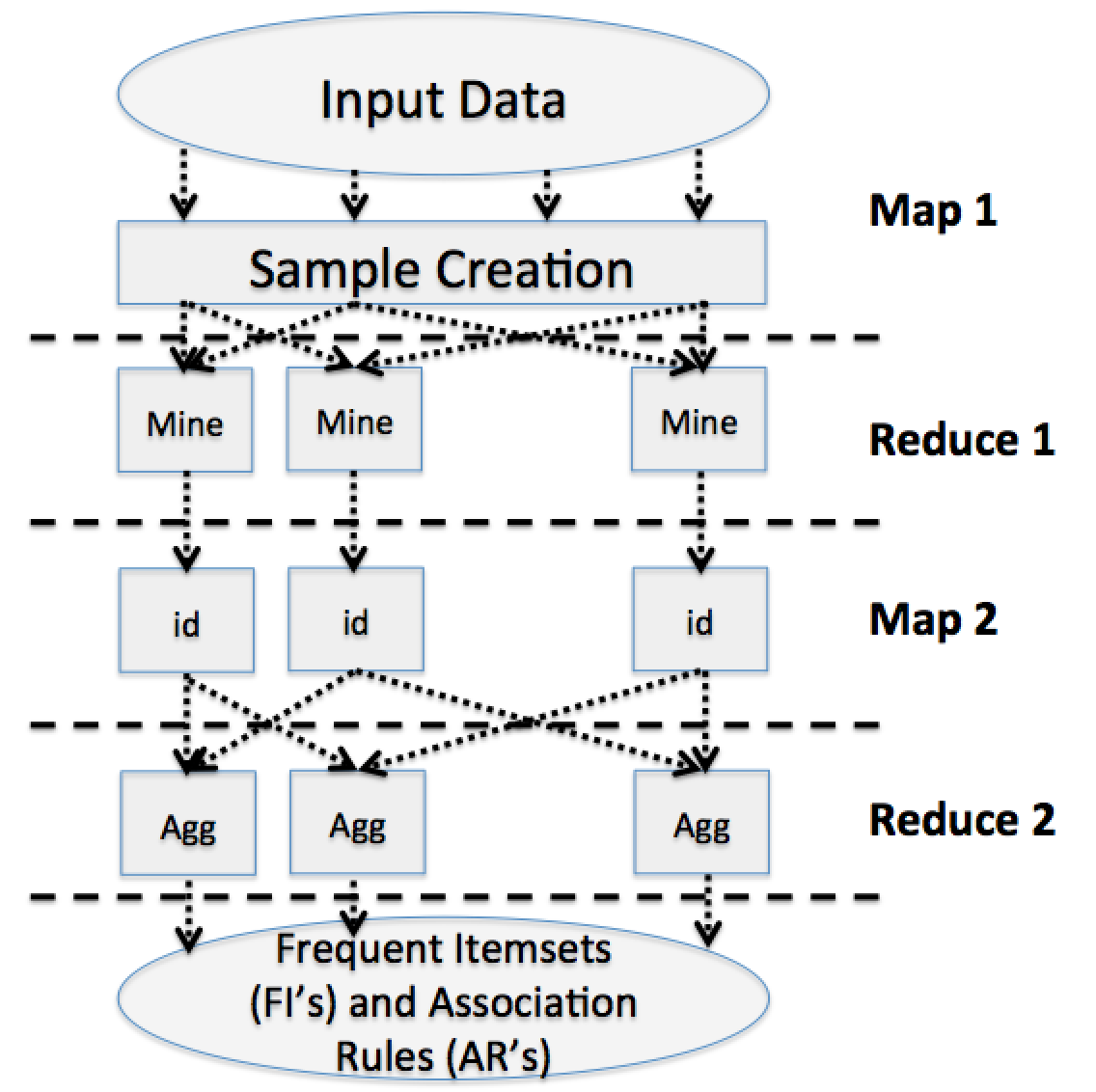}

\caption{An overview of the software architecture of PARMA.}
\label{parma}
\end{figure} 
As stressed in Figure \ref{parma}, PARMA creates multiple small random samples of the transactional dataset and runs a mining algorithm on the samples independently and in parallel.  
Results from each sample are aggregated and filtered to provide a single collection in output. 
The final result of PARMA is an approximation of the exact solution since it mines random subsets of the input dataset.

Table \ref{stateoftheartDMMLframeworks} presents the most popular data mining and machine learning techniques. 
For each technique, we list the programming model, the implemented techniques and the programming language. 
We notice that the input and the output of the above presented approaches are user-defined.

\begin{table*}[t]
\centering
\caption{Overview of data mining and machine learning techniques.}
\label{stateoftheartDMMLframeworks}
\scalebox{1}{
\begin{tabular}{|c|c|c|} \hline
\bfseries Approach& \bfseries Programming language & \bfseries Programming model \\ \hline 
NIMBLE  & JAVA  & MapReduce, OpenCL, MPI\\ \hline
Mahout  & JAVA & MapReduce \\ \hline
SystemML  & JAVA and DML & MapReduce \\ \hline
PARMA  & JAVA & MapReduce \\ \hline
\end{tabular}
}
\end{table*}

\section{Big graph analytics}
In this section, we first present some graph processing frameworks. Then, we highlight related works on big graph mining techniques. 
\subsection{Graph processing frameworks}
\subsubsection{Pregel}
\begin{figure}[h]
\centering

 \includegraphics[width=0.45\textwidth]{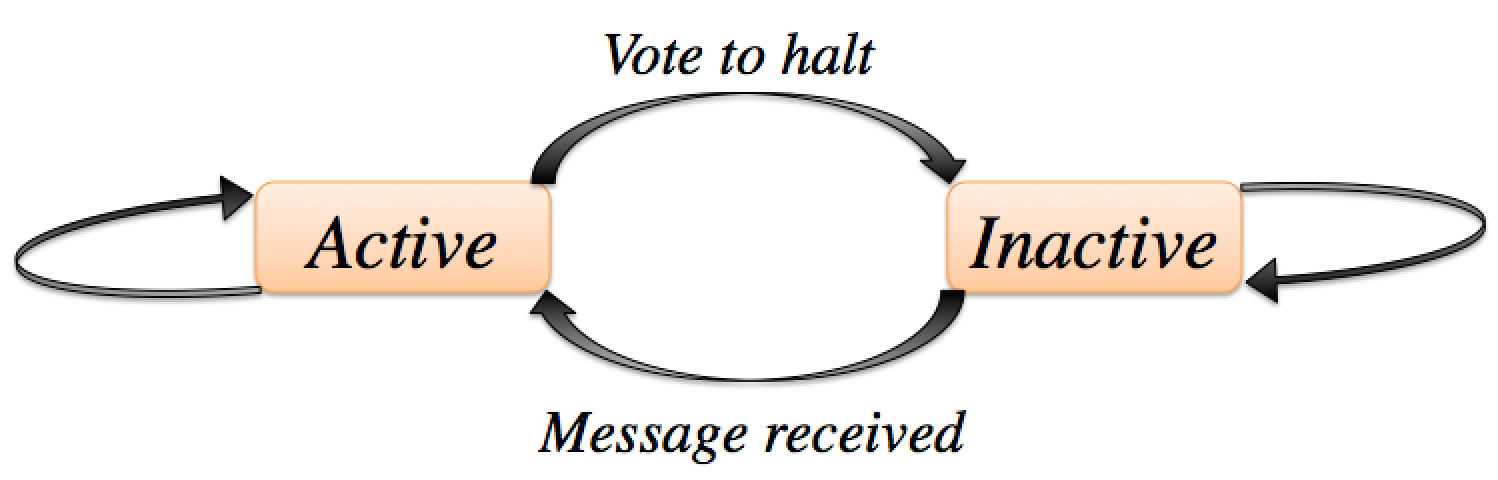}

\caption{Vertex's state machine in Pregel.}
\label{pregel}
\end{figure}

Pregel \cite{Malewicz_et_al} is a computational model suitable for large-scale graph processing problems. 
Pregel is vertex-centric model in which message exchanges occur among vertices of the input graph.  
The model has been designed for efficient, scalable, and fault-tolerant implementation on cluster of machines. 
Pregel's programming model is inspired by the Bulk Synchronous Parallel (BSP) model. 
Each vertex is associated to a state that controls its activity (see Figure \ref{pregel}). 
Each vertex can decide to halt its computation, but can be woken up at every point of the execution by an incoming message. 
At each superstep of the computation a user defined vertex program is executed for each active vertex. 
The user defined function will take the vertex and its incoming messages as input, change the vertex’s value and eventually send messages to other vertices through the outgoing edges. 
As defined by the BSP model, a synchronization barrier makes sure that all vertices have executed their instance of the user defined function. 
Pregel's implementation is not available to companies outside of Google, but Giraph \cite{Han:2015:GUB:2777598.2777604}, an open source implementation of the module was quickly introduced. 
Giraph offers a very similar API to pregel's. 
\subsubsection{Blogel}
Blogel \cite{Yan:2014:BBF:2733085.2733103} is a block-centric graph processing framework in which a block refers to a connected subgraph of the input graph, and message exchanges occur among blocks. 
It allows to execute both vertex-centric algorithms, block-centric algorithms and hybrid algorithms, in which all vertices execute before the entire block. 
These features allow Blogel to offer a very fast implementation of many graph algorithms. 
Blogel also offers readily available implementations of specialized partitioning algorithms, such as URL partitioning algorithms or 2D spatial partitioning, which can lead to much faster execution of the framework when additional data about vertices is available. 
Blogel operates in one of the following three computing modes, depending on the application: 
\begin{itemize}
 \item \textbf{B-mode.} In this mode, only block-level message exchanges are allowed. A job terminates when all blocks voted to halt and there is no pending message for the next superstep.
  \item \textbf{V-mode.} In this mode, only vertex-level message exchanges are allowed. A job terminates when all vertices voted to halt and there is no pending message for the next superstep.
   \item \textbf{VB-mode.} In this mode, V-mode is first activated and then B-mode is activated.  A job terminates only if all vertices and blocks voted to halt and there is no pending message for the next superstep. 
\end{itemize}
\subsubsection{GraphLab}
\begin{figure*}[t]

\centering
 \includegraphics[width=0.75\textwidth]{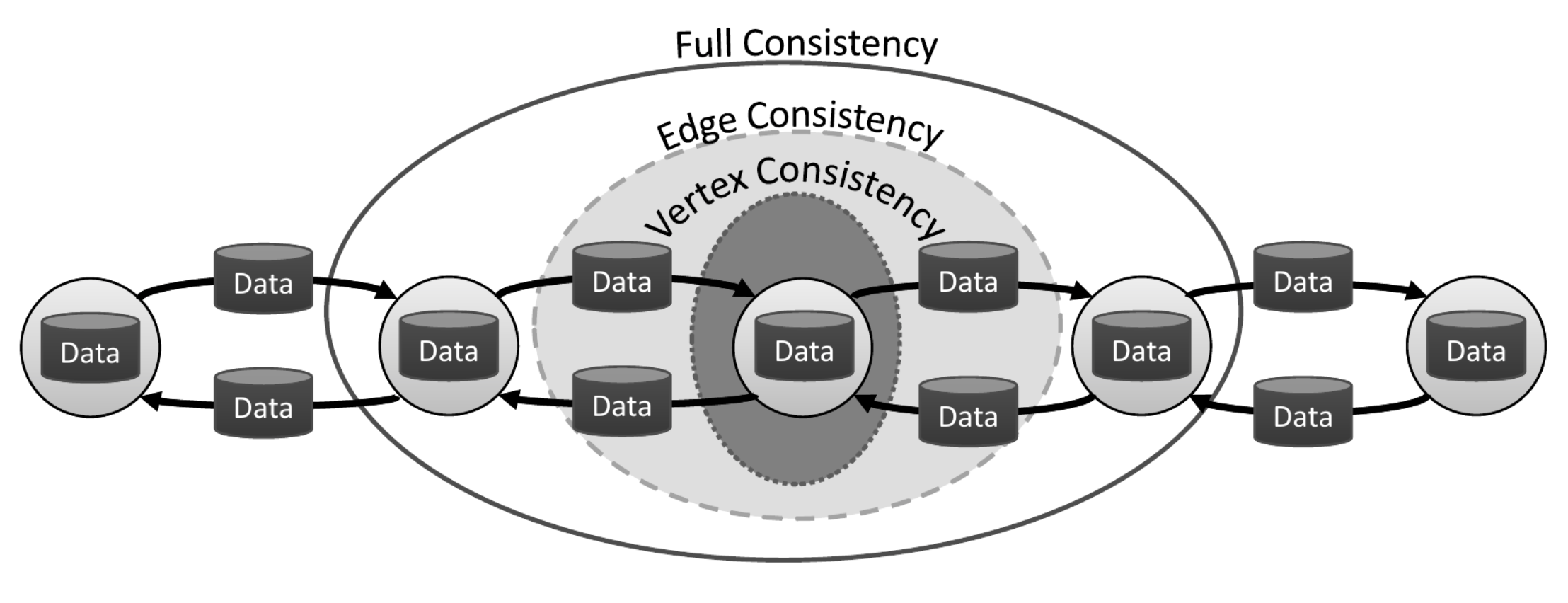}

\caption{View of the scope of a vertex in GraphLab.}
\label{scope}
\end{figure*}

GraphLab \cite{DBLP:journals/corr/LowGKBGH14} share the same motivation with Pregel. It has been designed to support large scale graph computation. 
While pregel targets Google’s large distributed system, GraphLab addresses shared memory parallel systems which means that there is more focus on parallel access of memory than on the issue of efficient message passing and synchronization. 
GraphLab's programming model is simple. Users must define an update function that, given as input a node and its entire neighborhood, can change all data associated to the scope of that node (its edges or its neighbors). 
Figure \ref{scope} shows the scope of a vertex: an update function called on that vertex will be able to read and write all data in its scope. 
It is important to notice that scopes can overlap, so simultaneously executing two update functions can result in a collision. 
Consequently, GraphLab offers three consistency models (a Fully Consistent, a Vertex Consistent or an Edge Consistent model), allowing users to trade off performance and consistency as appropriate for their computation. 

\subsubsection{PEGASUS}
PEGASUS \cite{kang:pegasus} is an open source graph mining library which performs typical graph mining tasks such as computing the diameter of a graph, computing the radius of each node and finding the connected components. 
The main idea of PEGASUS is the GIM-V primitive, standing for Generalized Iterative Matrix-Vector multiplication, which consists of a generalization of normal matrix-vector multiplication. 
PEGASUS customizes the GIM-V primitive and uses MapReduce in  order to handle with important large scale graph mining operations. It also provides several  optimizations such as block-multiplication and diagonal block iteration. 

\subsubsection{GraphX}
GraphX \cite{Xin:2013:GRD:2484425.2484427} is an Application Programming Interface (API) provided by Spark \cite{Zaharia:2012:RDD:2228298.2228301}, a generic distributed programming framework implemented as an extension of the mapreduce model. 
Spark introduces Resilient Distributed Datasets (RDD), that can be split in partitions and kept in memory by the machines of the cluster that is running the system. 
These RDD can be then passed to of predefined meta-functions such as map, reduce, filter or join, that will process them and return a new RDD. 
In GraphX, graphs are defined as a pair of two specialized RDD. The first one contains data related to vertices and the second one contains data related to edges of the graph. 
New operations are then defined on these RDD, to allow to map vertices's values via user defined functions, join them with the edge table or external RDDs, or also run iterative computation. \\ 

Table \ref{stateoftheartframeworks} presents the most popular graph processing frameworks. For each framework, we list the programming model, the type of resources used by the framework and whether the framework allows for asynchronous
execution or not.

\begin{table*}[t]
\centering
\caption{Overview of graph processing frameworks.}
\label{stateoftheartframeworks}
\scalebox{1}{
\begin{tabular}{|c|c|c|c|} \hline
\bfseries Framework & \bfseries Asynchronous execution& \bfseries Resources & \bfseries Programming model \\ \hline 
PEGASUS & No & Distributed system & Matrix operations\\ \hline
Pregel & No & Distributed system & Vertex-centric \\ \hline
Blogel & No & Distributed system & Graph-centric \\ \hline
GraphX & No & Distributed system & Edge-centric \\ \hline
GraphLab & Yes & Parallel systems & Vertex-centric \\ \hline
\end{tabular}
}
\end{table*}

\subsection{Pattern mining in big graphs}

In this section, we first present graph patterns and their associated tasks and applications.
Then, we present the frequent subgraph mining (FSM) task and we survey some recent FSM approaches in the context of big graphs.

\subsubsection{Applications of graph patterns}
Graph patterns aim to characterise complex graphs. They help finding properties that distinguish real-world graphs from random graphs and detect anomalies in a given graph.
Graph patterns are important for many applications such as chemoinformatics, bioinformatics and machine learning.

\begin{figure}[t]
 \centering
 \subfloat[A chemical compound]{\label{chem0}\includegraphics[scale=0.45]{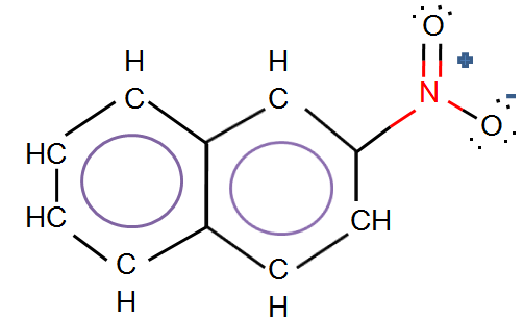}}
 
 \subfloat[A graph representation of a chemical compound]{\label{chem1}\includegraphics[scale=0.4]{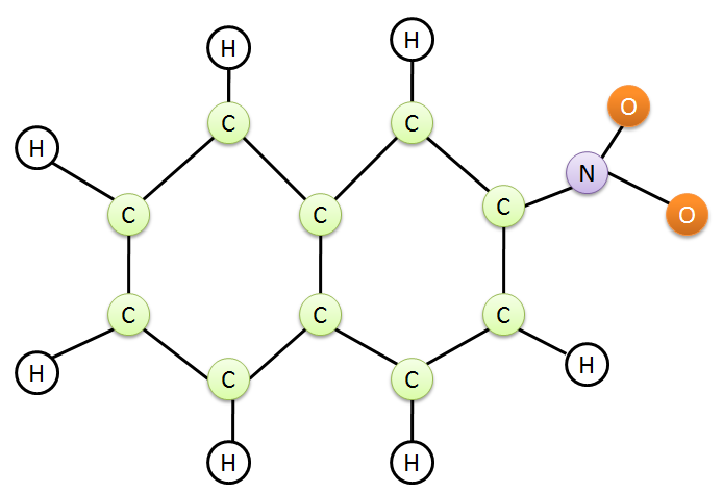}}
 \caption{Graph representation of a chemical compound.}
 \label{chemicalcompgraph}
\end{figure}

\textbf{Chemical patterns:} Chemical data is often represented as graphs in which the nodes correspond to atoms, and the links correspond to bonds between the atoms \cite{JSFA:JSFA2740490309,Ranu:2012:IMT:2247596.2247666,Wegner:2012:CHE:2366316.2366334}.
In some cases, chemical patterns may be used as individual nodes. In this case, the individual graphs are
quite small, though there are significant repetitions among the different nodes.
This leads to isomorphism challenges in applications such as graph matching. The isomorphism challenge is that the nodes in a given pair of graphs may match in a variety of ways.
The number of possible matches may be exponential in terms of the number of the nodes. 
Figure \ref{chemicalcompgraph} illustrates the graph representation of a real chemical compound.

\textbf{Biological patterns:}
 From a computer science point of view, the protein structure can be viewed as a set of elements. Each element can be an atom, an amino acid residue or a secondary structure fragment.
 Hence, several graph representations have been developed to preprocess protein structure,
 ranging from coarse representations in which each vertex is a secondary structure fragment \cite{Auron11011982,RNAPred} to fine representations in which each vertex is an atom \cite{DBLP:journals/bmcbi/SaidiMN10,saidi1}.
 Indeed, a protein interaction network can be represented by a graph where an edge links a couple of proteins when they participate in a particular biological function.
Biological patterns may correspond to important functional fragments in proteins such as active sites, feature positions and junction sites.


\textbf{Computer networked and Web data patterns:} In the case of computer networks and the web, the number of nodes in the underlying graph may be massive \cite{Erciyes:2013:DGA:2517714}. Since the number of nodes is massive, this can lead to a very large number of distinct edges. This is also referred to as the massive domain issue in networked data.
In such cases, the number of distinct edges may be so large, that it may be hard to hold in the available storage space.
Thus, techniques need to be designed to summarize and work with condensed representations of the graph data sets.
In some of these applications, the edges in the underlying graph may arrive in the form of a data stream.
In such cases, a second challenge arises from the fact that it may not be possible to store the incoming edges for future analysis.
Therefore, the summarization techniques are especially essential for this case. The stream summaries may be leveraged for future processing of the underlying graphs.

\begin{figure*}[t]
 \centering
 \includegraphics[scale=0.45]{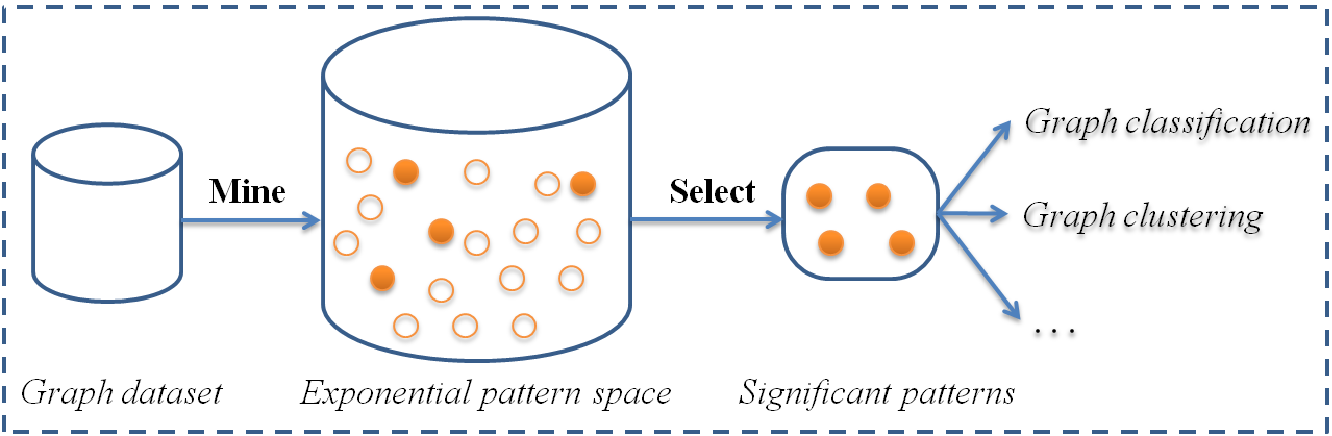}
 \caption{Graph patterns application pipeline.}
 \label{PatternApp}
\end{figure*}

Figure \ref{PatternApp} depicts the pipeline of graph applications built on frequent patterns.

In this pipeline, frequent patterns are mined first; then significant patterns are selected based on user-defined objective functions for different applications.

\subsubsection{FSM problem formulation}
There are two separate problem formulations for FSM: (1) graph transaction based FSM and (2) single graph based FSM.
In graph transaction based FSM, the input data comprises a collection of medium-size graphs called transactions.
In single graph based FSM the input data, as the name implies, comprise one very large graph.

Let $DB$ be a graph database. Each graph $G = (V, E)$ of $DB$, is given as a collection of nodes $V$ and edges $E$. 
We denote by $|V|$ the number of nodes of $G$ and by $|E|$ the number of edges of $G$ (also called graph size). 
If two nodes $u \in V$ and $v \in V$ and $\{u, v\} \in E$ then $u$ and $v$ are said to be adjacent nodes. 

\begin{definition}[Graph]
A graph is denoted as $G = (V,E)$, where:
\begin{itemize}
\item $V$ is a set of nodes (vertices).
\item $E \subseteq   V \times V$ is a set of edges (links).
\end{itemize}
\end{definition}


\begin{definition}[Graph isomorphism] 
An isomorphism of graphs $G$ and $H$ is a bijection $f:V(G) \longrightarrow V(H)$
 such that any two vertices $u$ and $v$ of $G$ are adjacent in $G$ if and only if $f(u)$ and $f(v)$ are adjacent in $H$.
\end{definition}

\begin{definition}[Subgraph]
A graph $G'=(V', E')$ is a subgraph of another graph $G=(V, E)$ iff:
 \begin{itemize}
 \item $V' \subseteq V$, and
 \item $E' \subseteq E \cap (V' \times V')$.
\end{itemize}
\end{definition}

\begin{definition}[Subgraph isomorphism]
A graph $G'(V',E')$ is subgraph-isomorphic to a graph $G(V,E)$ if there exists an injective function $f:V'(G') \longrightarrow V(G)$ such that $(f(u),f(v)) \in E$ holds for each $(u,v) \in E'$.
The function $f$ represents an embedding of $G'$ in $G$, $G'$ is called subgraph of $G$, and $G$ is called supergraph of $G'$.
\end{definition}

The definitions of subgraph support, graph transaction based FSM and single graph based FSM are given as follows.

\begin{definition}[Subgraph relative support] 
Given a graph database $DB = \{G_1, \ldots, G_K\}$, the \emph{relative support} of a subgraph $G'$ is defined by
\begin{equation}
Support(G', DB)=\frac{\sum_{i=1}^{k} \sigma(G',G_i)}{k},
\end{equation}
where
\[\sigma(G',G_i)= \begin{cases} 1, & \mbox{if } G' \mbox{ has a subgraph isomorphism with } G_i, \\ 0, & \mbox{otherwise. } \end{cases} \]

\end{definition}

\begin{definition}[Graph transaction based FSM]
Given a minimum support threshold $\theta \in [0, 1]$, the frequent subgraph mining task with respect to $\theta$ is finding all subgraphs with a support greater than $\theta$, i.e., the set $SG(DB,\theta) = \{(A,Support(A,DB)): A$ is a subgraph of $DB$ and $Support(A,DB) \ge \theta \}$.
\end{definition}

In single graph based FSM, the most intuitive way to measure the support of a subgraph $G'$ in a graph $G$ is to count its isomorphisms. 

\subsubsection{Properties of FSM techniques}
\label{sectionFSMProp}
Frequent subgraph mining approaches perform differently in order to mine frequent subgraphs from a graph dataset.
Such differences are related to the search strategy, the generation of candidate patterns strategy and the support computing method.

\textbf{Search strategy}
There are two basic search strategies employed for mining frequent subgraphs \cite{JiangCZ13,Sujata2013}: the depth first search (DFS) strategy and the breadth first search (BFS) strategy. 
The DFS strategy is a method for traversing or searching tree or graph data structures.
It starts at the root (selecting a node as the root in the graph case) and explores as far as possible along each branch before backtracking.
%
The BFS strategy is limited to essentially two operations:
\begin{enumerate}
\item Visit and inspect a node of a graph,
\item Gain access to visit the nodes that neighbor the currently visited node.
\end{enumerate}
The BFS begins at a root node and inspects all the neighboring nodes.
Then for each of those neighbor nodes in turn, it inspects their neighbor nodes which were unvisited, and so on.

\begin{figure}[h]
\centering
\subfloat[Apriori-based approach ]{\label{cost12}\includegraphics[scale=0.37]{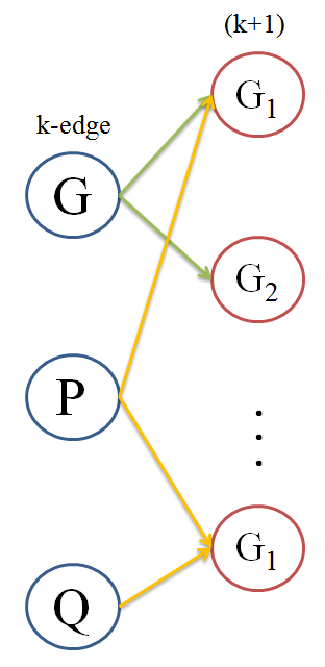}}
\subfloat[Pattern growth-based approach]{\label{cost12}\includegraphics[scale=0.37]{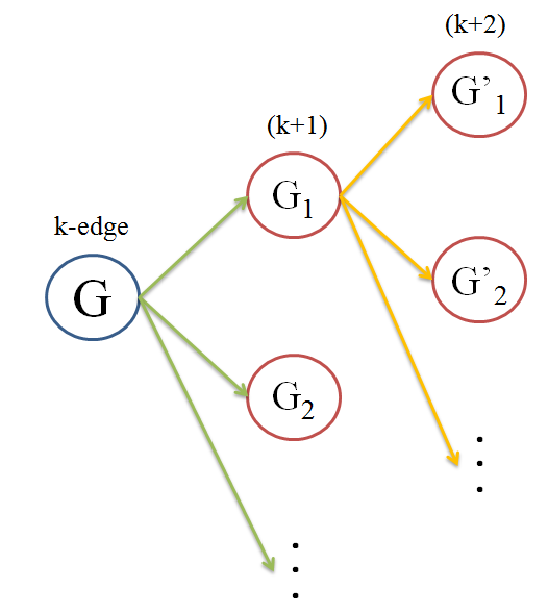}}
\caption{Apriori-based vs pattern growth-based approach.}
\label{aprvgrow}
\end{figure}

\textbf{Generation of candidate patterns} The generation of candidate patterns is the core element of the frequent subgraph discovery process.
Here, we consider two types of approaches: Apriori-based and pattern growth-based. Figure \ref{aprvgrow} shows the difference between the two approaches.

Apriori-based approaches share similar characteristics with Apriori-based frequent itemset mining algorithms \cite{Agrawal:1994:FAM:645920.672836}.
The search for frequent graphs starts with graphs of small size, and proceeds in a bottom-up manner.
At each iteration, the size of the newly discovered frequent substructures is increased by one (see Figure \ref{cost12}).
These new substructures are first generated by joining two similar but slightly different frequent subgraphs that were discovered already.
The frequency of the newly formed graphs is then checked.
The Apriori-based algorithms have considerable overhead when two size-k frequent substructures are joined to generate size-(k+1) graph candidates.
Typical Apriori-based frequent substructure mining algorithms are discussed in the following paragraphs.

The pattern-growth mining algorithm extends a frequent graph by adding a new edge, in every possible position as shown in Figure \ref{cost12}.
A potential problem with the edge extension is that the same graph can be discovered many times.

\textbf{Support computing} Several methods are used for graph counting.
Some frequent subgraph mining algorithms use transaction identifier (TID) lists for frequency counting. Each frequent subgraph has a list of transaction identifiers which support it. For computing frequency of a $k$ subgraph, the intersection of the TID lists of $(k - 1)$ subgraphs is computed. Also, DFS lexicographic ordering can be used for frequency evaluation. Here, each graph is mapped into a DFS sequence followed by construction of a lexicographic order among them based on these sequences, and thus a search tree is developed. The minimum DFS code obtained from this tree for a particular graph is the canonical label of that graph which helps in evaluating the frequency.
Embedding lists are used for support computing. For all graphs, a list is stored of embedding tuples that consist of (1) an index of an embedding tuple in the embedding list of the predecessor graph and (2) the identifier of a graph in the database and a node in that graph.
The frequency of a structure is determined from the number of different graphs in its embedding list. Embedding lists are quick, but they do not scale very well to large databases.
The other approach is based on maintaining a set of active graphs in which occurrences are repeatedly recomputed.

\textbf{Types of patterns}
Several kinds of subgraph patterns can be mined with existing frequent subgraph mining algorithms.
\begin{itemize}
 \item \textbf{Frequent subgraphs} A frequent subgraph is a subgraph whose support is no less than a minimum support threshold. Algorithms including gSpan \cite{gspan} and FFSM \cite{ffsm} aim to mine frequent subgraphs. 
  \item \textbf{Closed frequent subgraphs} A set of closed subgraph patterns has the same expressive power as the full set of subgraph patterns under the same minimum support threshold, 
  because the latter can be generated by the derived set of closed graph patterns. CloseGraph method \cite{Yan:2003:CMC:956750.956784} is a well known algorithm for mining closed frequent subgraphs. 

  \item \textbf{Maximal frequent subgraphs} The maximal pattern set is a subset of the closed pattern set. It is usually more compact than the closed pattern set. However, we cannot use it to reconstruct the entire set of frequent patterns. 
  Although the set of closed or maximal subgraphs is much smaller than the set of frequent ones, real-world graphs contain an exponential number of subgraphs. Algorithms like SPIN \cite{Huan04spin:mining} and MARGIN \cite{Thomas:2010:MMF:1839490.1839491} aim to mine maximal frequent subgraphs. 

\end{itemize}

\subsubsection{FSM techniques in big graphs}
With the exponential growth in both the graph size and the number of graphs in databases, several distributed solutions have been proposed for pattern mining on a single large graphs and on massive graph databases. 
Specifically, these works focus of subgraph patterns and describe frequent subgraph mining (FSM) approaches for big graphs \cite{Liu:2009:MPF:1616625.1616658} \cite{Aridhi2015213} \cite{Luo:2011:TES:1996686.1996690} \cite{Hill:2012:IMA:2382936.2383055}. 

\textbf{MRPF.} In \cite{Liu:2009:MPF:1616625.1616658}, the authors propose the MRPF algorithm for finding patterns from a complex and large network. 
\begin{figure}[t]
\centering

 \includegraphics[width=8cm,height=10cm]{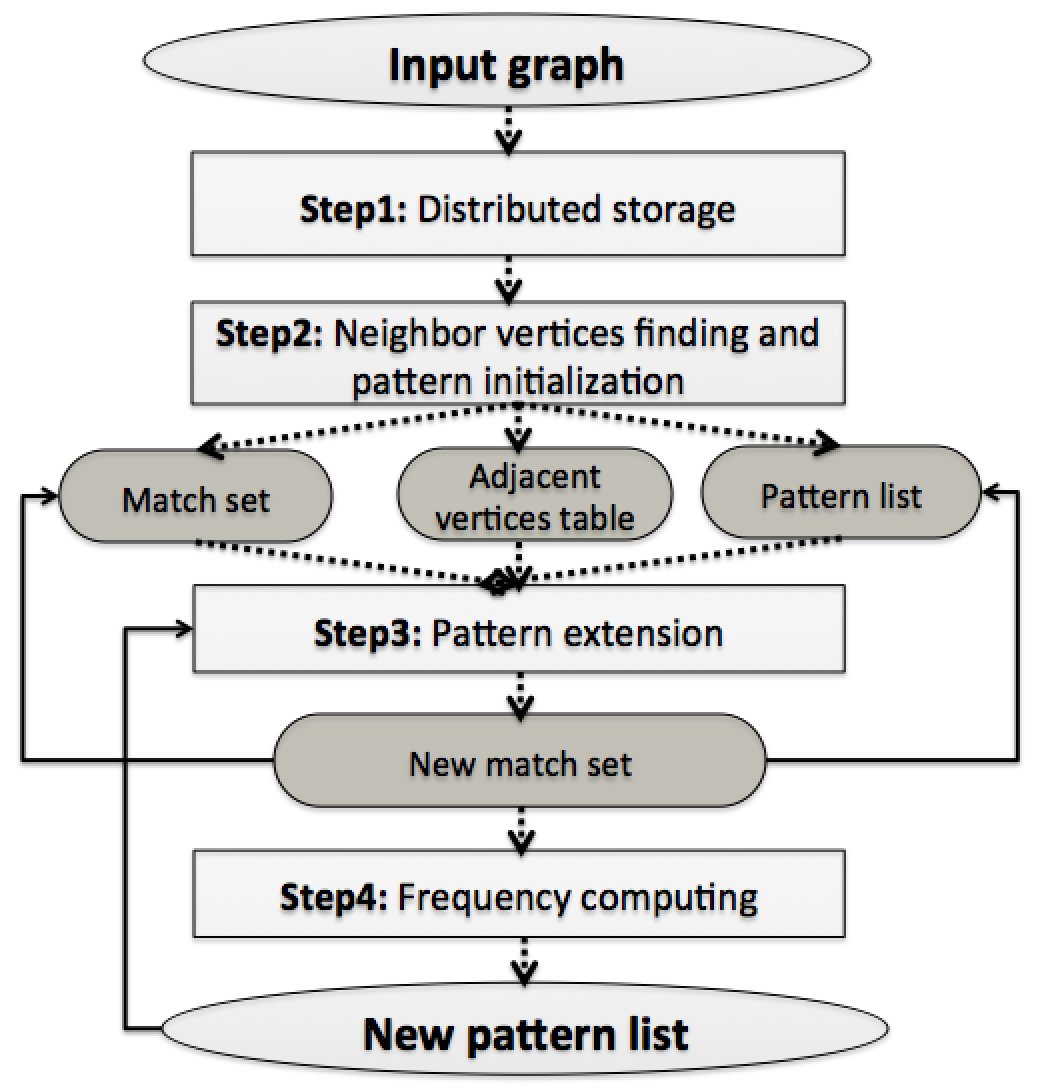}

\caption{An overview of the software architecture of MRPF.}
\label{mrpf}
\end{figure} 
As illustrated in Figure \ref{mrpf}, the algorithm is divided into four steps: (1) distributed storage of the graph, (2) neighbor vertices finding and pattern initialization, (3) pattern extension, and (4) frequency computing. 
Each step is implemented by a MapReduce pass. In each MapReduce pass, the task is divided into a number of sub-tasks of the same size and each sub-task is distributed to a node of the cluster.
MRPF uses an extended mode to find the target size pattern.
That is trying to add one more vertex to the matches of $i$-size patterns to create patterns of size $i+1$.
The extension does not stop until patterns reach the target size.
The proposed algorithm is applied to prescription network in order to find some commonly used prescription network motifs that provide the possibility to discover the law of prescription compatibility.

\textbf{Hill etal.'s approach.} The work presented in \cite{Hill:2012:IMA:2382936.2383055} presents an iterative MapReduce-based approach for frequent subgraph mining. 
\begin{figure}[t]
\centering \includegraphics[width=7cm,height=10cm]{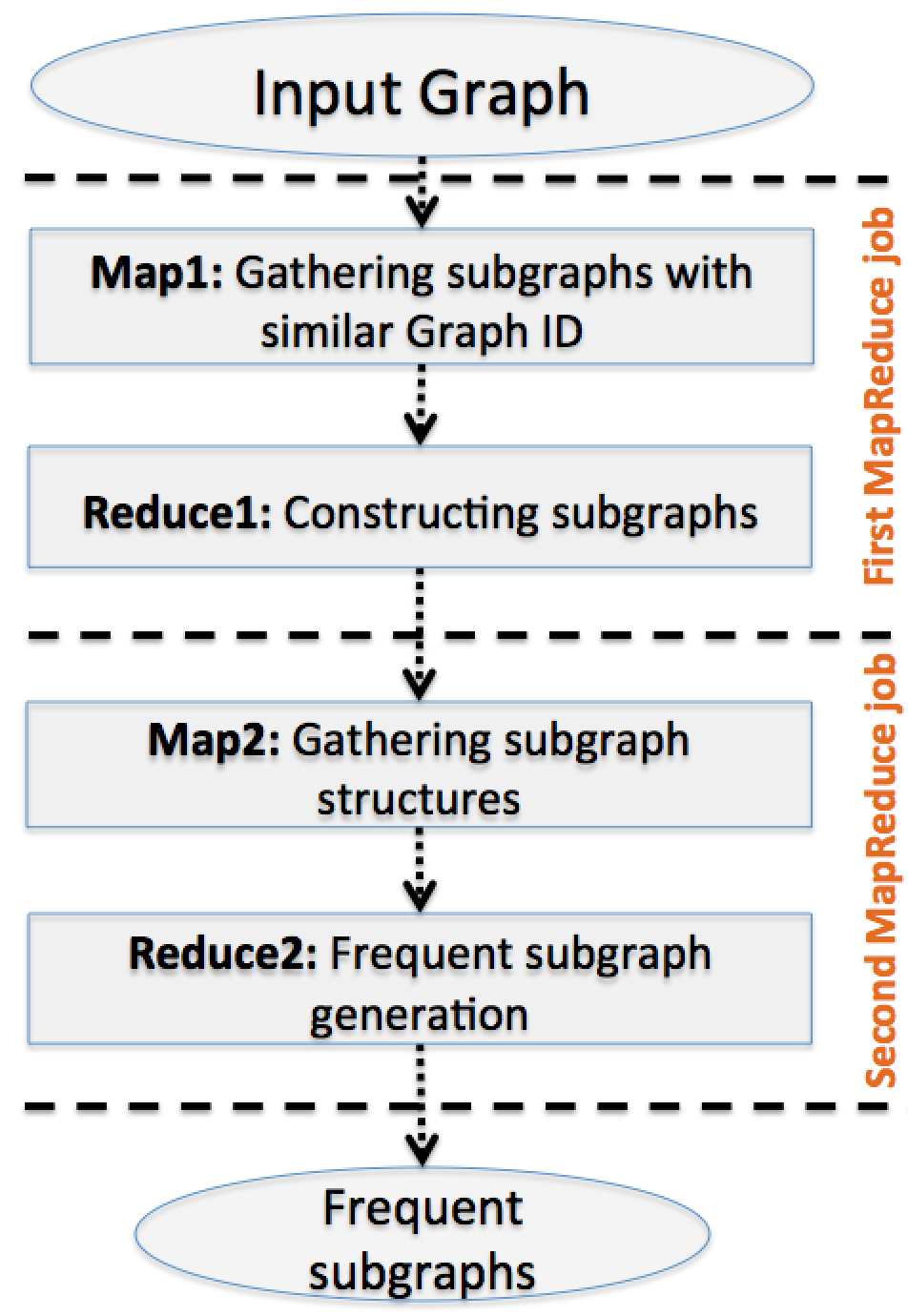}

\caption{An overview of the system overview of Hill etal.'s approach.}
\label{hill}
\end{figure} 
As stressed in Figure \ref{hill}, Hill etal.'s approach generates the set of frequent subgraphs by performing two heterogeneous MapReduce jobs per iteration: (1) gathering subgraphs for the construction of the next generation of subgraphs, and (2) counting these structures to remove irrelevant data.
The first MapReduce job aims to construct the next generation of subgraphs. 
Its associated Map function sends the subgraph to the correct reducer using the graph identifier as a key. 
All the subgraphs of size $k-1$ with the same graph identifier are gathered for the Reduce function. 
Single edges in these subgraphs are used to generate the next generation of possible subgraphs of size $k$. 
The subgraph is encoded as a string. All labels alphabetized are kept and the special markers are used to designate different nodes with the same labels. 
The results of this step are subgraphs of size $k$ and graph identifiers. 
The second MapReduce job aims to output the frequent subgraphs. 
The map function of this job has the responsibility of taking in the encoded strings representing subgraphs of size-$k$ and corresponding graph identifiers as well as outputting the subgraph as a key and the node identification numbers and graph identifiers as values. 
The reduce function gathers (per iteration) on label only subgraph structures. 
The main task of the reduce function is to compute the support of these subgraphs. 
The label markers are removed at this point.
The outputs of iteration $k$ are all subgraphs of size $k$ that meet the given user defined support.


\textbf{Aridhi etal.'s approach.} In \cite{Aridhi2015213}, the authors propose a novel approach for large-scale subgraph mining, using the MapReduce framework. 
\begin{figure}[t]
\centering
 \includegraphics[width=8.5cm,height=8cm]{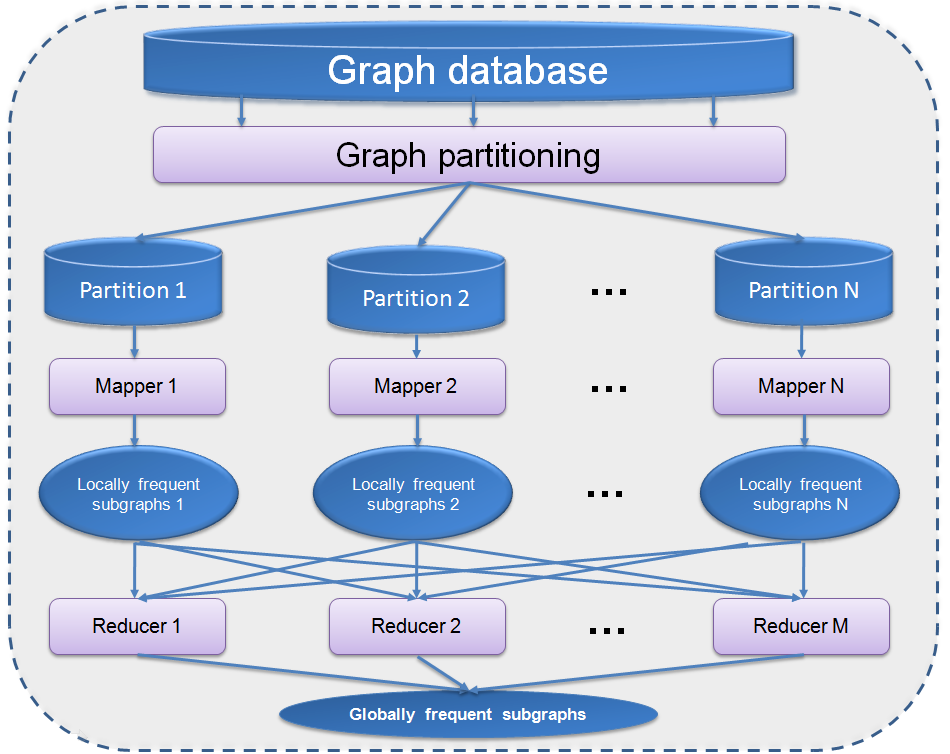}

\caption{An overview of the software architecture of Aridhi et al.'s approach.}
\label{aridhi}
\end{figure} 

As shown in Figure \ref{aridhi}, Aridhi et al's approach works as follows:
\begin{enumerate}
\item Input graph database is partitioned into N partitions. Each partition will be processed by a mapper machine.  
\item Mapper i reads the assigned data partition and generates the corresponding locally frequent subgraphs, i.e., frequent subgraphs inside each partition.
\item The reducer uses the sets of locally frequent subgraphs as input and computes for each subgraph its support in the whole graph database. 
Then, it outputs the set of globally frequent subgraphs, i.e., subgraphs that are frequent in the whole graph database. 
\end{enumerate}
The proposed approach provides a data partitioning technique that consider data characteristics \cite{Aridhi2015213}. 
It uses the densities of graphs in order to partition the input data. Such a partitioning technique allows a balanced computational loads over the distributed collection of machines and replace the default arbitrary partitioning technique of MapReduce. 

\textbf{Luo etal.'s approach.} In \cite{Luo:2011:TES:1996686.1996690}, the authors propose an approach to subgraph search over a graph database under the MapReduce framework.
The main idea of the proposed approach is first to build inverted edge indexes for graphs in the database, and then to retrieve data only related to the query subgraph by using the built indexes to answer the query.

\begin{figure}[t]
\centering
 \includegraphics[width=8.5cm,height=8cm]{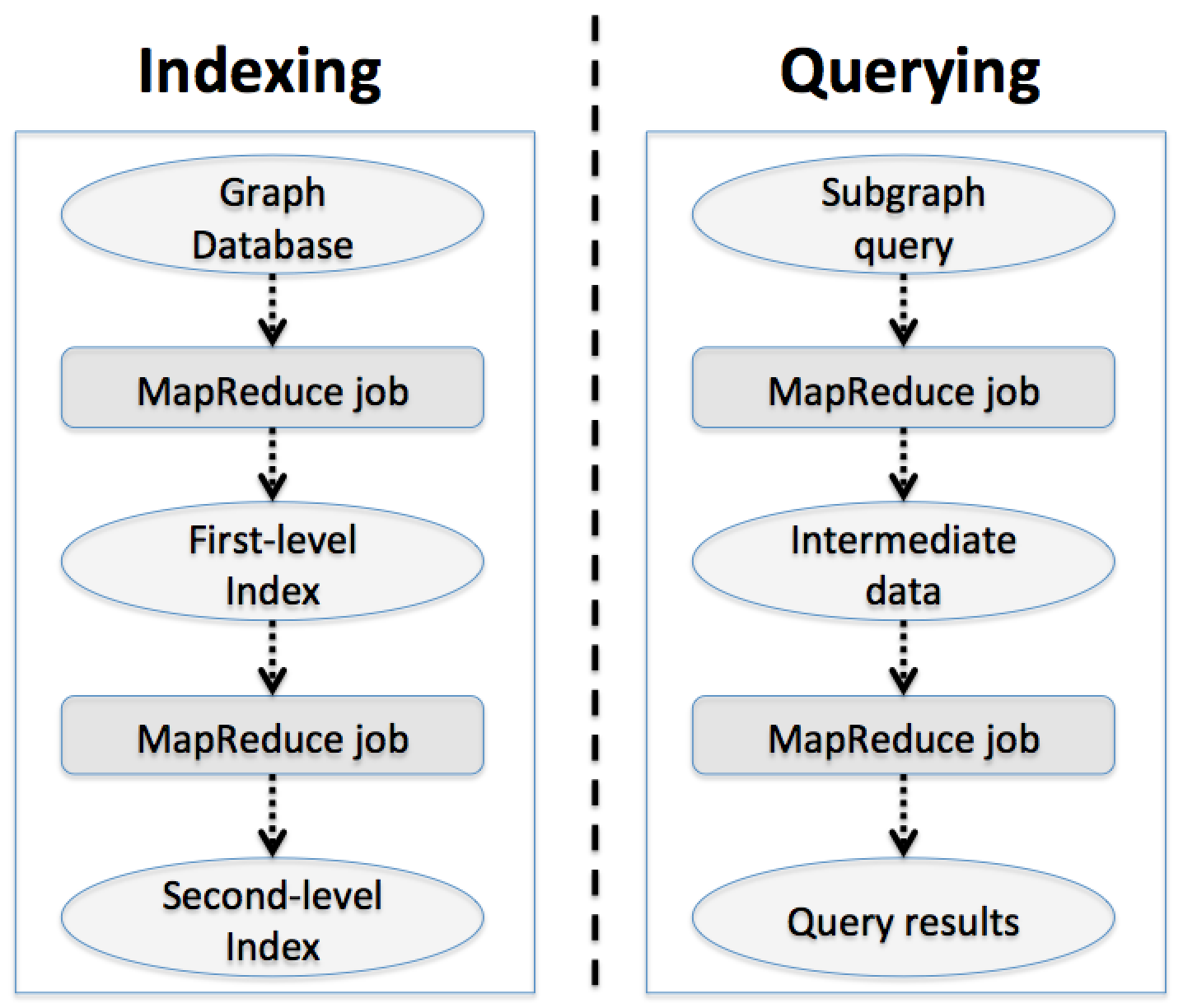}

\caption{An overview of Luo et al.'s approach.}
\label{luo}
\end{figure} 
As shown in Figure \ref{luo}, Luo etal.'s approach performs two MapReduce jobs to build inverted indexes. The first MapReduce job is responsible for building inverted indexes for each unique edge in the graph database, the second one aims to build indexes over the inverted indexes for each unique edge built in the first phase. 
In order to process queries, two MapReduce jobs are launched. The first MapReduce job is to retrieve the candidate results by using indexing information, the second one is to evaluate the final query results by employing set intersection operations. 




Table \ref{stateoftheart} presents the most popular approaches of distributed graph mining techniques. 
It describes the input, the output of each approach and indicates the used programming model. 

\begin{table*}[t]
\centering
\caption{Summary of popular pattern mining techniques in big graphs. }
\label{stateoftheart}
\scalebox{1}{
\begin{tabular}{|c|c|c|c|} \hline
\bfseries Approach & \bfseries Input & \bfseries Output & \bfseries Programming model \\ \hline 
Aridhi etal.'s approach \cite{Aridhi2015213} & Graph database & Frequent subgraphs & MapReduce\\ \hline
PARMA \cite{parma} & Transactional database & Frequent itemsets & MapReduce\\ \hline
HADI \cite{HADI}& Graph database &  Diameter of each graph & MapReduce  \\ \hline
Zhao etal.'s approach \cite{Zhao:2007:IPE:1769115.1769128} & Graph database &  Eigenvalue of each graph & MPI/OpenMPI \\ \hline
MRPF \cite{Liu:2009:MPF:1616625.1616658} & Single graph + subgraph model&  Frequent subgraphs & MapReduce \\ \hline
Luo etal.'s approach \cite{Luo:2011:TES:1996686.1996690} & A graph database&  Frequent subgraphs & MapReduce \\ \hline
Hill etal.'s approach \cite{Hill:2012:IMA:2382936.2383055} & A graph database + subgraph model & Frequent subgraphs & MapReduce \\ \hline
\end{tabular}
}
\end{table*}

\subsubsection{Global graph pattern mining in big graphs}

Global graph patterns are similar in spirit to the characteristic measures used in descriptive statistics (like mean, variance and skew). 
In the work of \cite{Cohen:2009:GTM:1591877.1592018}, the authors give an investigation into the feasibility of decomposing useful graph operations into a series of MapReduce processes. 
Such a decomposition could enable implementing the graph algorithms on a cloud, in a streaming environment, or on a single computer.
In \cite{HADI}, the authors propose HADI algorithm, a solution for mining diameter in massive graphs on the top of MapReduce. 
HADI have been used to analyze the largest public web graph, with billions of nodes and edges.
In Zhao etal.'s approach \cite{Zhao:2007:IPE:1769115.1769128}, the authors propose solutions for computing eigenvalue in massive graphs. 

\section{Conclusion}
In this work, we presented a survey on frameworks and techniques for Big Data analytics with a focus on graph data. 
We first presented large scale data mining and machine learning techniques. 
Then, we presented graph processing frameworks and current pattern mining techniques in big graphs. 
With reference to the literature several graph processing frameworks have been proposed and many pattern mining approaches have been proposed with 
respect to several types of input data, to produce many different kinds of patterns. 
We have also adopted a categorization of both distributed data mining and machine learning techniques, graph processing frameworks and large scale pattern mining approaches. 
Data mining and machine learning techniques were described according to their related programming model and the used programming language. 
Graph processing frameworks were described according to their related programming model, the type of resources used by each framework and whether the framework allows asynchronous execution or not. 
Pattern mining approaches were described according to the input, the output of each approach and the used programming model. 

\bibliographystyle{abbrv}

\bibliography{biblio}

\begin{thebibliography}{10}

\bibitem{Agrawal:1994:FAM:645920.672836}
R.~Agrawal and R.~Srikant.
\newblock Fast algorithms for mining association rules in large databases.
\newblock In {\em Proceedings of the 20th International Conference on Very
  Large Data Bases}, VLDB '94, pages 487--499, San Francisco, CA, USA, 1994.
  Morgan Kaufmann Publishers Inc.

\bibitem{AlJarrah201587}
O.~Y. Al-Jarrah, P.~D. Yoo, S.~Muhaidat, G.~K. Karagiannidis, and K.~Taha.
\newblock Efficient machine learning for big data: A review.
\newblock {\em Big Data Research}, 2(3):87 -- 93, 2015.

\bibitem{Alvarez}
J.~I. Alvarez-Hamelin, A.~Barrat, A.~Vespignani, and et~al.
\newblock k-core decomposition of internet graphs: hierarchies, self-similarity
  and measurement biases.
\newblock {\em Networks and Heterogeneous Media}, 3(2):371, 2008.

\bibitem{Aridhi2015213}
S.~Aridhi, L.~d'Orazio, M.~Maddouri, and E.~Mephu~Nguifo.
\newblock Density-based data partitioning strategy to approximate large-scale
  subgraph mining.
\newblock {\em Information Systems}, 48:213 -- 223, 2015.

\bibitem{Auron11011982}
P.~E. Auron, W.~P. Rindone, C.~P. Vary, J.~J. Celentano, and J.~N. Vournakis.
\newblock Computer-aided prediction of rna secondary structures.
\newblock {\em Nucleic Acids Research}, 10(1):403--419, 1982.

\bibitem{Bahmani:2012:DSS:2140436.2140442}
B.~Bahmani, R.~Kumar, and S.~Vassilvitskii.
\newblock Densest subgraph in streaming and mapreduce.
\newblock {\em Proc. VLDB Endow.}, 5(5):454--465, Jan. 2012.

\bibitem{Cohen:2009:GTM:1591877.1592018}
J.~Cohen.
\newblock Graph twiddling in a mapreduce world.
\newblock {\em Computing in Science and Engg.}, 11(4):29--41, July 2009.

\bibitem{mapreduce}
J.~Dean and S.~Ghemawat.
\newblock {MapReduce: simplified data processing on large clusters}.
\newblock {\em Commun. ACM}, 51(1):107--113, Jan. 2008.

\bibitem{Erciyes:2013:DGA:2517714}
K.~Erciyes.
\newblock {\em Distributed Graph Algorithms for Computer Networks}.
\newblock Springer Publishing Company, Incorporated, 2013.

\bibitem{mahout}
A.~S. Foundation, I.~Drost, T.~Dunning, J.~Eastman, O.~Gospodnetic,
  G.~Ingersoll, J.~Mannix, S.~Owen, and K.~Wettin.
\newblock Apache mahout, 2010.

\bibitem{hbasebook}
N.~Garg.
\newblock {\em HBase Essentials}.
\newblock Packt Publishing, 2014.

\bibitem{nimble}
A.~Ghoting, P.~Kambadur, E.~Pednault, and R.~Kannan.
\newblock Nimble: a toolkit for the implementation of parallel data mining and
  machine learning algorithms on mapreduce.
\newblock In {\em Proceedings of the 17th ACM SIGKDD international conference
  on Knowledge discovery and data mining}, KDD '11, pages 334--342, New York,
  NY, USA, 2011. ACM.

\bibitem{Ghoting:2011:SDM:2004686.2005625}
A.~Ghoting, R.~Krishnamurthy, E.~Pednault, B.~Reinwald, V.~Sindhwani,
  S.~Tatikonda, Y.~Tian, and S.~Vaithyanathan.
\newblock Systemml: Declarative machine learning on mapreduce.
\newblock In {\em Proceedings of the 2011 IEEE 27th International Conference on
  Data Engineering}, ICDE '11, pages 231--242, Washington, DC, USA, 2011. IEEE
  Computer Society.

\bibitem{5992567}
C.~Giatsidis, D.~Thilikos, and M.~Vazirgiannis.
\newblock Evaluating cooperation in communities with the k-core structure.
\newblock In {\em Advances in Social Networks Analysis and Mining (ASONAM),
  2011 International Conference on}, pages 87--93, July 2011.

\bibitem{Han:2015:GUB:2777598.2777604}
M.~Han and K.~Daudjee.
\newblock Giraph unchained: Barrierless asynchronous parallel execution in
  pregel-like graph processing systems.
\newblock {\em Proc. VLDB Endow.}, 8(9):950--961, May 2015.

\bibitem{Hill:2012:IMA:2382936.2383055}
S.~Hill, B.~Srichandan, and R.~Sunderraman.
\newblock An iterative mapreduce approach to frequent subgraph mining in
  biological datasets.
\newblock In {\em Proceedings of the ACM Conference on Bioinformatics,
  Computational Biology and Biomedicine}, BCB '12, pages 661--666, New York,
  NY, USA, 2012. ACM.

\bibitem{ffsm}
J.~Huan, W.~Wang, and J.~Prins.
\newblock Efficient mining of frequent subgraphs in the presence of
  isomorphism.
\newblock In {\em Proceedings of the Third IEEE International Conference on
  Data Mining}, ICDM '03, pages 549--, Washington, DC, USA, 2003. IEEE Computer
  Society.

\bibitem{Huan04spin:mining}
J.~Huan, W.~Wang, and J.~Prins.
\newblock Spin: Mining maximal frequent subgraphs from graph databases.
\newblock In {\em In KDD}, pages 581--586, 2004.

\bibitem{JiangCZ13}
C.~Jiang, F.~Coenen, and M.~Zito.
\newblock A survey of frequent subgraph mining algorithms.
\newblock {\em Knowledge Eng. Review}, 28(1):75--105, 2013.

\bibitem{Jin201559}
X.~Jin, B.~W. Wah, X.~Cheng, and Y.~Wang.
\newblock Significance and challenges of big data research.
\newblock {\em Big Data Research}, 2(2):59 -- 64, 2015.

\bibitem{HADI}
U.~Kang, C.~Tsourakakis, A.~P. Appel, C.~Faloutsos, and J.~Leskovec.
\newblock Hadi: Fast diameter estimation and mining in massive graphs with
  hadoop, 2008.

\bibitem{kang:pegasus}
U.~Kang, C.~E. Tsourakakis, and C.~Faloutsos.
\newblock {PEGASUS: A Peta-Scale Graph Mining System Implementation and
  Observations}.
\newblock In {\em {Proceedings of the 2009 Ninth IEEE International Conference
  on Data Mining}}, {ICDM '09}, pages 229--238, Washington, DC, USA, 2009. IEEE
  Computer Society.

\bibitem{Liu:2009:MPF:1616625.1616658}
Y.~Liu, X.~Jiang, H.~Chen, J.~Ma, and X.~Zhang.
\newblock Mapreduce-based pattern finding algorithm applied in motif detection
  for prescription compatibility network.
\newblock In {\em Proceedings of the 8th International Symposium on Advanced
  Parallel Processing Technologies}, APPT '09, pages 341--355, Berlin,
  Heidelberg, 2009. Springer-Verlag.

\bibitem{DBLP:journals/corr/LowGKBGH14}
Y.~Low, J.~E. Gonzalez, A.~Kyrola, D.~Bickson, C.~Guestrin, and J.~M.
  Hellerstein.
\newblock Graphlab: {A} new framework for parallel machine learning.
\newblock {\em CoRR}, abs/1408.2041, 2014.

\bibitem{Luo:2011:TES:1996686.1996690}
Y.~Luo, J.~Guan, and S.~Zhou.
\newblock Towards efficient subgraph search in cloud computing environments.
\newblock In {\em Proceedings of the 16th international conference on Database
  systems for advanced applications}, DASFAA'11, pages 2--13, Berlin,
  Heidelberg, 2011. Springer-Verlag.

\bibitem{Malewicz_et_al}
G.~Malewicz, M.~H. Austern, A.~J. Bik, J.~C. Dehnert, I.~Horn, N.~Leiser, and
  G.~Czajkowski.
\newblock {Pregel: a system for large-scale graph processing}.
\newblock In {\em {Proceedings of the 2010 international conference on
  Management of data}}, {SIGMOD '10}, pages 135--146, New York, NY, USA, 2010.
  ACM.

\bibitem{JSFA:JSFA2740490309}
Y.~Miyashita, M.~Ishikawa, and S.-I. Sasaki.
\newblock Classification of brandies by pattern recognition of chemical data.
\newblock {\em Journal of the Science of Food and Agriculture}, 49(3):325--333,
  1989.

\bibitem{pig}
C.~Olston, B.~Reed, U.~Srivastava, R.~Kumar, and A.~Tomkins.
\newblock Pig latin: A not-so-foreign language for data processing.
\newblock In {\em Proceedings of the 2008 ACM SIGMOD International Conference
  on Management of Data}, SIGMOD '08, pages 1099--1110, New York, NY, USA,
  2008. ACM.

\bibitem{JTLU23}
R.~Patuelli, A.~Reggiani, P.~Nijkamp, and F.-J. Bade.
\newblock The evolution of the commuting network in germany: Spatial and
  connectivity patterns.
\newblock {\em Journal of Transport and Land Use}, 2(3), 2010.

\bibitem{Ranu:2012:IMT:2247596.2247666}
S.~Ranu and A.~K. Singh.
\newblock Indexing and mining topological patterns for drug discovery.
\newblock In {\em Proceedings of the 15th International Conference on Extending
  Database Technology}, EDBT '12, pages 562--565, New York, NY, USA, 2012. ACM.

\bibitem{parma}
M.~Riondato, J.~A. DeBrabant, R.~Fonseca, and E.~Upfal.
\newblock Parma: a parallel randomized algorithm for approximate association
  rules mining in mapreduce.
\newblock In {\em Proceedings of the 21st ACM international conference on
  Information and knowledge management}, CIKM '12, pages 85--94, New York, NY,
  USA, 2012. ACM.

\bibitem{saidi1}
R.~Saidi, S.~Aridhi, E.~{Mephu Nguifo}, and M.~Maddouri.
\newblock {Feature extraction in protein sequences classification: a new
  stability measure}.
\newblock In {\em {Proceedings of the ACM Conference on Bioinformatics,
  Computational Biology and Biomedicine}}, {BCB '12}, pages 683--689, New York,
  NY, USA, 2012. ACM.

\bibitem{DBLP:journals/bmcbi/SaidiMN10}
R.~Saidi, M.~Maddouri, and E.~{Mephu Nguifo}.
\newblock {Protein sequences classification by means of feature extraction with
  substitution matrices}.
\newblock {\em BMC Bioinformatics}, 11:175, 2010.

\bibitem{Seidman1983269}
S.~B. Seidman.
\newblock Network structure and minimum degree.
\newblock {\em Social Networks}, 5(3):269 -- 287, 1983.

\bibitem{Sujata2013}
S.~J. Suryawanshi and S.~M. Kamalapur.
\newblock Algorithms for frequent subgraph mining.
\newblock {\em International Journal of Advanced Research in Computer and
  Communication Engineering}, 2(3):1545--1548, 2013.

\bibitem{Tatikonda_et_al}
S.~Tatikonda and S.~Parthasarathy.
\newblock {Mining tree-structured data on multicore systems}.
\newblock {\em Proc. VLDB Endow.}, 2(1):694--705, Aug. 2009.

\bibitem{Thomas:2010:MMF:1839490.1839491}
L.~T. Thomas, S.~R. Valluri, and K.~Karlapalem.
\newblock Margin: Maximal frequent subgraph mining.
\newblock {\em ACM Trans. Knowl. Discov. Data}, 4(3):10:1--10:42, Oct. 2010.

\bibitem{hive}
A.~Thusoo, J.~S. Sarma, N.~Jain, Z.~Shao, P.~Chakka, S.~Anthony, H.~Liu,
  P.~Wyckoff, and R.~Murthy.
\newblock Hive: A warehousing solution over a map-reduce framework.
\newblock {\em Proc. VLDB Endow.}, 2(2):1626--1629, Aug. 2009.

\bibitem{Wegner:2012:CHE:2366316.2366334}
J.~K. Wegner, A.~Sterling, R.~Guha, A.~Bender, J.-L. Faulon, J.~Hastings,
  N.~O'Boyle, J.~Overington, H.~Van~Vlijmen, and E.~Willighagen.
\newblock Cheminformatics.
\newblock {\em Commun. ACM}, 55(11):65--75, Nov. 2012.

\bibitem{Worlein:2005:QCS:2101235.2101276}
M.~W\"{o}rlein, T.~Meinl, I.~Fischer, and M.~Philippsen.
\newblock A quantitative comparison of the subgraph miners mofa, gspan, ffsm,
  and gaston.
\newblock In {\em Proceedings of the 9th European conference on Principles and
  Practice of Knowledge Discovery in Databases}, PKDD'05, pages 392--403,
  Berlin, Heidelberg, 2005. Springer-Verlag.

\bibitem{Xin:2013:GRD:2484425.2484427}
R.~S. Xin, J.~E. Gonzalez, M.~J. Franklin, and I.~Stoica.
\newblock Graphx: A resilient distributed graph system on spark.
\newblock In {\em First International Workshop on Graph Data Management
  Experiences and Systems}, GRADES '13, pages 2:1--2:6, New York, NY, USA,
  2013. ACM.

\bibitem{Yan:2014:BBF:2733085.2733103}
D.~Yan, J.~Cheng, Y.~Lu, and W.~Ng.
\newblock Blogel: A block-centric framework for distributed computation on
  real-world graphs.
\newblock {\em Proc. VLDB Endow.}, 7(14):1981--1992, Oct. 2014.

\bibitem{gspan}
X.~Yan and J.~Han.
\newblock {gSpan: Graph-Based Substructure Pattern Mining}.
\newblock In {\em {Proceedings of the 2002 IEEE International Conference on
  Data Mining}}, {ICDM '02}, pages 721--724, Washington, DC, USA, 2002. IEEE
  Computer Society.

\bibitem{Yan:2003:CMC:956750.956784}
X.~Yan and J.~Han.
\newblock Closegraph: mining closed frequent graph patterns.
\newblock In {\em Proceedings of the ninth ACM SIGKDD international conference
  on Knowledge discovery and data mining}, KDD '03, pages 286--295, New York,
  NY, USA, 2003. ACM.

\bibitem{Zaharia:2012:RDD:2228298.2228301}
M.~Zaharia, M.~Chowdhury, T.~Das, A.~Dave, J.~Ma, M.~McCauley, M.~J. Franklin,
  S.~Shenker, and I.~Stoica.
\newblock Resilient distributed datasets: A fault-tolerant abstraction for
  in-memory cluster computing.
\newblock In {\em Proceedings of the 9th USENIX Conference on Networked Systems
  Design and Implementation}, NSDI'12, pages 2--2, Berkeley, CA, USA, 2012.
  USENIX Association.

\bibitem{Zhao:2007:IPE:1769115.1769128}
Y.~Zhao, X.~Chi, and Q.~Cheng.
\newblock An implementation of parallel eigenvalue computation using dual-level
  hybrid parallelism.
\newblock In {\em Proceedings of the 7th international conference on Algorithms
  and architectures for parallel processing}, ICA3PP'07, pages 107--119,
  Berlin, Heidelberg, 2007. Springer-Verlag.

\bibitem{RNAPred}
M.~Zuker and D.~Sankoff.
\newblock Rna secondary structures and their prediction.
\newblock {\em Bulletin of Mathematical Biology}, 46(4):591--621, 1984.

\end{thebibliography}

\end{document}